\documentclass[aps,prl,superscriptaddress,twocolumn,showpacs,titlepage=false]{revtex4-1}
\usepackage{graphicx}
\usepackage{amsmath}
\usepackage{amsmath}
\usepackage{mathrsfs}
\usepackage{dsfont}
\usepackage{amssymb}
\usepackage{subfigure}
\usepackage{color}
\usepackage{blkarray}
\usepackage[breaklinks=true,colorlinks=true,linkcolor=blue,urlcolor=blue,citecolor=blue]{hyperref}

\def\ket#1{{\left| #1 \right\rangle}}

\definecolor{amber}{rgb}{1.0, 0.49, 0.0}
\usepackage[dvipsnames]{xcolor}

\usepackage{verbatim}
\usepackage{amsthm}
\usepackage{enumerate}
\usepackage{bbm}
\usepackage{hyperref}

\usepackage{verbatim}
\usepackage{graphicx}
\usepackage{amssymb}
\usepackage{amsmath}
\usepackage{amsthm}
\usepackage{enumerate}
\usepackage{bbm}
\usepackage{hyperref}

\newcommand*{\cE}{\mathcal{E}}

\newcommand*{\Tr}{\mathrm{Tr}}

\newcommand{\beq}{\begin{equation}}
\newcommand{\eeq}{\end{equation}}

\begin{document}

\title{Decoherence, quantum Darwinism, and the generic emergence of our objective classical reality}
\author{Paul A. Knott}
\email{Paul.Knott@nottingham.ac.uk}
\affiliation{Centre for the Mathematics and Theoretical Physics of Quantum Non-Equilibrium Systems (CQNE), School of Mathematical Sciences, University of Nottingham, University Park, Nottingham NG7 2RD, UK}
%\author{Tommaso Tufarelli}
%\affiliation{Centre for the Mathematics and Theoretical Physics of Quantum Non-Equilibrium Systems (CQNE), School of Mathematical Sciences, University of Nottingham, University Park, Nottingham NG7 2RD, UK}
%\author{Marco Piani}
%\affiliation{SUPA and Department of Physics, University of Strathclyde, Glasgow, G4 0NG, UK}
%\author{Gerardo Adesso}
%\affiliation{Centre for the Mathematics and Theoretical Physics of Quantum Non-Equilibrium Systems (CQNE), School of Mathematical Sciences, University of Nottingham, University Park, Nottingham NG7 2RD, UK}
\date{\today}

%================Abstract===================%
\begin{abstract}
In this article I aim to provide an intuitive and non-technical introduction to decoherence and quantum Darwinism. Together these theories explain how our classical reality emerges from an underlying quantum mechanical description. Here I focus on two aspects of this and explain, firstly, how decoherence can tell us why we never see macroscopic superpositions, such as dead-and-alive cats, in our classical surroundings; and secondly I introduce and then provide a resolution to the so-called preferred basis problem. I then introduce a remarkable recent result by Brand{\~a}o et al. [Nat. Commun. 6, 7908 (2015)], which show that certain aspects of classicality are generic phenomena that emerge from the basic mathematical structure of quantum mechanics. This is in stark contrast to the majority of previous results in this field that focused on specific models that cannot realistically be scaled up to explicitly answer questions about the macroscopic world. Finally, I demonstrate how decoherence and quantum Darwinism can shed significant light on the measurement problem, and I discuss the implications for how we should interpret quantum mechanics.
%How does our objective classical reality emerge from an underlying quantum mechanical substrate? The common resolution to this used to be that the classical world must obey different rules to the quantum world, or that quantum mechanics is not complete and there must be an additional mechanism that prevents macroscopic superposition states from persisting. However, in recent decades the theories of decoherence and quantum Darwinism have demonstrated that, perhaps surprisingly, the Schrödinger equation alone can explain the emergence of our objective classical reality. Despite the successes, results in this field generally relied on studying specific models, but this changed with a quite remarkable result by Brand{\~a}o et al. [Nat. Commun. 6, 7908 (2015)] who showed that certain aspects of objectivity are generic phenomena that emerge in a model-independent way from the basic mathematical structure of quantum mechanics. In this article I give an intuitive introduction to decoherence and quantum Darwinism, focusing in particular on the preferred basis problem. I then introduce the results of Brand{\~a}o et al., and explain how their theorems imply the generic emergence of objectivity. Finally, I demonstrate how decoherence can shed significant light on the measurement problem, and I discuss the implications for how we should interpret quantum mechanics.
\end{abstract}
\maketitle

Since the early development of quantum mechanics, there has always been an apparent divide between the quantum and classical worlds. Quantum mechanical objects perpetually exist in superposition states: electrons orbiting nuclei live in a delocalised state with an undefined position; and photons can be said to be in all possible locations simultaneously until they are detected. Yet at the other end of the scale, macroscopic objects such as cats and dogs are never found to be in a superposition. The equally mysterious properties of entanglement and nonlocal correlations are likewise never witnessed on a day-to-day basis. Despite this, macroscopic objects are made of quantum particles, and this raises a question that has always sparked debate and controversy: Can quantum mechanics alone describe the macroscopic classical world, or do we need an additional or extended theory? If the former, then numerous phenomena need to be explained using only the theory of quantum mechanics, such as why we never see macroscopic objects in a superposition, and less obvious questions such as why macroscopic objects exhibit \emph{objective} properties.

%===========Fig:================
%\begin{figure}[t!]
%\centering
%\includegraphics[width=9.5cm]{ART.jpeg}
%\caption{In Quantum Darwinism, the objective classical reality (image of a cat) emerges from an underlying quantum mechanical description  (bottom layer, illustrating superposition effects).}
%\label{artwork0}
%\end{figure}
%====================================

One focus of this article is the notion of \emph{objectivity}. A property is said to be objective if multiple observers agree on the details of that property. For example, all of the objects around you have an objective position. If I ask the question ``where is my cup of tea?'', then my answer will be the same as any other observer looking at the tea (at least to a good approximation). So why is it, and how is it, that properties of the objects we see around us are objective? The answer to this question is not obvious when we think about the quantum mechanical properties of the constituents of these macroscopic objects. Despite this, we will see in this article that quantum mechanics \emph{does} lead to objectivity. I will show this by introducing the powerful framework of decoherence -- which itself explains why macroscopic superpositions do not persist -- and an extension of decoherence known as quantum Darwinism \cite{ZurekRMP,SchlosshauerRMP,ZurekQD,Ollivier2004,Ollivier2005,Blume2006,Horodecki2015,le2018strong,korbicz2017generic}. But before this, to fully specify the notion of objectivity we are concerned with, I will first introduce the ``preferred basis problem''.

\subsection{The preferred basis problem}
A particularly interesting problem -- which we will see is closely linked with objectivity -- is the question of why there is a “preferred basis” in the macroscopic world. To understand what is meant by this, imagine the spin of an electron, which can be in two states, $\left|\uparrow \right\rangle$ or $\left|\downarrow \right\rangle$, as represented by the basis $\{\left|\uparrow \right\rangle, \left|\downarrow \right\rangle\} $. If the state of the electron is $\left|\uparrow \right\rangle$, then if we measure the electron in the basis $\{\left|\uparrow \right\rangle, \left|\downarrow \right\rangle\} $, then we know that with certainty we will always obtain the measurement outcome ``up''. However, if we prepare the electron in a balanced superposition of $\left|\uparrow \right\rangle$ and $\left|\downarrow \right\rangle$, given by
\beq
{1 \over \sqrt{2}} ( \left|\uparrow \right\rangle + \left|\downarrow \right\rangle ),
\eeq
then the probability of measuring up will be $1/2$, and similarly the probability of measuring down will be $1/2$. But there are other bases, which are equally valid, that can be used to measure the electron. Take the basis corresponding to the electron being in the state $\left|\rightarrow \right\rangle$ or $\left|\leftarrow \right\rangle$, which are defined as 
\begin{align} 
\left|\rightarrow\right\rangle &= {1 \over \sqrt{2}} ( \left|\uparrow \right\rangle + \left|\downarrow \right\rangle )\\
\left|\leftarrow\right\rangle &= {1 \over \sqrt{2}} (  \left|\uparrow \right\rangle - \left|\downarrow \right\rangle ).
\end{align}
If we measure the electron in this basis, then if the state of the electron is $\left|\uparrow \right\rangle$, then the probability of measuring left is $1/2$, which is the same probability as measuring right. But if the electron is in the state $ \left|\rightarrow\right\rangle $, then clearly the probability of measuring right is $1$.
This example will be familiar to many quantum theory researchers, and it is often taken for granted that there are different bases with which we can measure in. Indeed, in the lab the state $\left|\uparrow \right\rangle$ can be changed to the state $ \left|\rightarrow\right\rangle $ by simply rotating the electron $90 $ degrees; or by rotating the lab $90$ degrees; or even just by redefining which direction is up! There is nothing special about any basis -- they are all equally valid; there is no ``preferred basis'' here.

In our everyday experiences of the macroscopic world, the notion of measuring in a particular basis is very different than in the example of an electron above. To introduce why this is so, I will use the example of Schr\"dinger's cat: a cat is placed in a box, isolated from its surroundings, with a device that contains a radioactive atom and a vial of poison. If the atom decays, then the device is designed to release the poison, killing the cat. What if the atom is prepared in a superposition state of decaying and not decaying? The Schr\"{o}dinger equation -- the equation governing the evolution of isolated systems in quantum mechanics -- predicts that if the atom is in a superposition state, then this will lead to the cat being in a superposition state:
\beq
{1 \over \sqrt{2}} ( \left| dead \right\rangle+\left| alive \right\rangle )
\eeq 
where $\left| dead \right\rangle$ represents the dead cat, and likewise for alive. (Note that the state $\left| dead \right\rangle$ represents the dead cat, and also the decayed radioactive atom and the released poison. Similarly for alive.) If the box is then opened, then the obvious question to ask for whoever opens the box is: ``is the cat dead or alive?''. This corresponds to performing a measurement in the basis $\{\left| dead\right\rangle,\left| alive\right\rangle\} $. But if we follow the same prescription as in the case of the electron, then there are different bases we can consider, such as $\{\left| +\right\rangle,\left| -\right\rangle\} $, where we define:
\begin{align} 
\left| + \right\rangle &= {1 \over \sqrt{2}} ( \left| dead \right\rangle + \left| alive\right\rangle )\\
\left| -\right\rangle &= {1 \over \sqrt{2}} ( \left| dead\right\rangle - \left| alive\right\rangle).
\end{align}
It is intuitively clear from our experience that the cat is never in a superposition state such as $\left| +\right\rangle$ or $\left| -\right\rangle$. But how can we be sure of this? If we could measure in the basis $\{\left| +\right\rangle,\left| -\right\rangle\} $ then by obtaining the measurement outcome plus we could determine with high confidence that the cat is in the state $\left| +\right\rangle$. But we never perform measurements in the basis $\{\left| +\right\rangle,\left| -\right\rangle\} $. Why is this? We can extend this idea to any measurement of the \emph{position} of objects around us. We invariably asked the question ``where is this object, here or there?'', but we never ask the question ``is it in a superposition `here $+$ there' or `here $-$ there'?''. (Of course, in general we perform position measurements with multiple possible outcomes, not just two, but the same argument holds true in the more general case.) 

It seems that nature restricts our measurements to always be in some particular basis (this basis is termed the \emph{pointer basis}). In fact we seem to be forced, but without knowing it, to measure in this basis. This article will address this particular issue, and I will explain why, and under what conditions, objects that are described by quantum mechanics can only be measured in this particular basis.

This fits into the idea of \emph{objectivity} in the following way: if two different observers measure in a different basis, then the notion of objectivity doesn't even have a clear meaning. Put another way, if two different people ask different questions, it wouldn't make sense to say that they agree on the answers. In the example above, if two different observers are asked to peer into the box containing Schr\"{o}dinger's cat and describe what they see, there are two things we can be certain of: i) the observers will measure in the same basis, and ii) they will agree on the outcomes of their measurements. It is instructive -- and entertaining -- to imagine how human society would function in a world in which there \emph{wasn't} a preferred basis!

To introduce some important terminology, roughly speaking the notion of \emph{which basis to measure in} can be slightly generalised to the notion of \emph{which observable to measure}. But for the purpose of this article, the \emph{measurement basis} and the \emph{observable} can be considered the same thing, and I will interchange these terms throughout.

\subsection{Decoherence}
To explain how objectivity and a preferred basis emerge in quantum mechanics, I must first introduce the frameworks of decoherence and quantum Darwinism. Decoherence rests on the important observation that quantum systems are rarely isolated, but invariably interact with an inaccessible environment. By studying how the system and environment interact, I will show how decoherence can be used to explain why certain superposition states are fragile. In turn we will see why macroscopic superposition states -- such as a cat being dead and alive -- are never seen in the real world.

%===========Fig:================
\begin{figure}[t!]
\centering
\includegraphics[width=8cm]{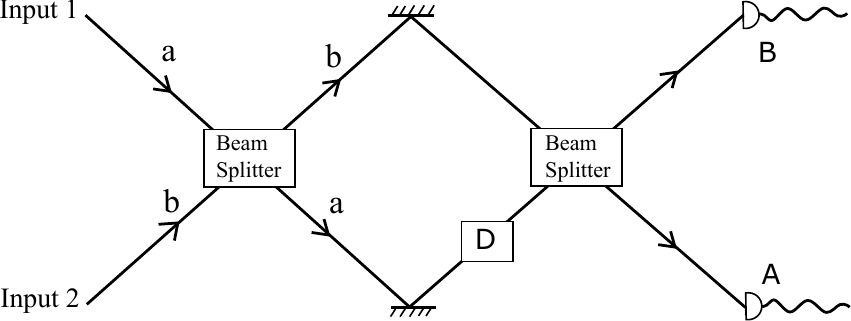}
\caption{A Mach-Zehnder interferometer.}
\label{intf}
\end{figure}
%====================================

I will begin with an intuitive, but perhaps unconventional, introduction to decoherence. Imagine inputting a single photon into a two path interferometer, as shown in Figure~\ref{intf}. The state in which there is a single photon in path $a$ and a vacuum in path $b$ is represented by the state $\ket{1,0}_{ab} $, where the subscript corresponds to the paths shown in the Figure. The photon enters the first beam splitter, which results in the following transformation:
\begin{equation}\label{bs1}
\ket{1,0}_{ab} \rightarrow {1 \over \sqrt{2}} ( \ket{1,0}_{ab} + \ket{0,1}_{ab} )
\end{equation}
The beam splitter has created a superposition state: the photon is in a superposition of being in path $a$ and path $b$. But how do we \emph{know} it is in a superposition? This may seem like a simple question, but it is crucial to understanding quantum mechanics. The answer is that the only way we can determine that something is in a superposition is using the phenomenon of interference.

(Note: the statement that interference is the only way to confirm a superposition assumes that our measurements are restricted: in this example we are assuming we can only measure whether the photon is in the upper or lower detector -- we make the perfectly reasonable assumption that we can't measure whether the photon is in a superposition of these detectors. Similar reasonable assumptions about what we can measure follow throughout this article.)

The interferometer provides a natural way to demonstrate interference (hence the name), as I now show. What happens to the state in equation~(\ref{bs1}) when it goes through the second beam splitter? Because quantum mechanics is linear, we can work out separately what happens to each part of the superposition state of the photon. We know what happens to the state $\ket{1,0}_{ab} $, and it can be shown that the state $\ket{0,1}_{ab} $ is transformed in the following way:
\begin{equation}
\ket{0,1}_{ab} \rightarrow {1 \over \sqrt{2}} ( \ket{0,1}_{ab} - \ket{1,0}_{ab} ). 
\end{equation}
The minus sign is necessary to keep the transformation unitary. Putting this together, the superposition state is transformed as follows:
\begin{align}
{1 \over \sqrt{2}} ( \ket{1,0}_{ab} &+ \ket{0,1}_{ab} ) \notag\\
&\rightarrow {1 \over {2}} ( \ket{1,0}_{ab} + \ket{0,1}_{ab} + \ket{0,1}_{ab} - \ket{1,0}_{ab} ). \notag\\
&= \ket{0,1}_{ab}
\end{align}
This is the archetypal demonstration of destructive interference: the state $\ket{1,0}_{ab} $ cancels with itself, leaving only the state $\ket{0,1}_{ab} $. If we place detectors at the outputs of the interferometer, we will see that, no matter how many times we repeat the experiment, we will \emph{never} observe a photon at detector $A$.

Now imagine we place a small detector on path $a$, as shown by the box labelled $D$ in Figure~\ref{intf}. For reasons that will become clear later, I will refer to this detector as the ``environment''. The environment starts in the state $r$ (where $r$ stands for the environment being ``ready''). The environment does not disturb the path of the photon, but if the photon is in path $a$ then it will record this by changing state accordingly:
\begin{align}
&\ket{1,0, r} \rightarrow \ket{1,0, a} 
&\ket{0,1, r} \rightarrow \ket{0,1, b} 
\end{align}
Therefore, the total state of the photon (in between the beam splitters) interacts with the environment as follows:
\begin{align}
{1 \over \sqrt{2}} ( \ket{1,0,r} + &\ket{0,1,r} ) \notag \\
&\rightarrow {1 \over \sqrt{2}} ( \ket{1,0, a} + \ket{0,1, b} ) \label{afterD}
\end{align}
We can see that the environment and the photon are now in an entangled state. We can now pass the photon through the final beam splitter, but the result is quite different from before:
\begin{align}
{1 \over \sqrt{2}} ( \ket{1,0, a} &+ \ket{0,1, b} ) \\ &\rightarrow {1 \over {2}} ( \ket{1,0, a} + \ket{0,1, a} + \ket{0,1, b} - \ket{1,0, b} ) \notag
\end{align}
We now make an assumption that is crucial to decoherence: we say that the two states of the environment, $a$ and $b$, are orthogonal to one another (i.e. $\langle a | b \rangle=0$; I will return to this notion of orthogonal environmental states later). Because of this orthogonality, the equation above demonstrates that there has been no destructive interference. The fact that the environment has recorded the state of the photon \emph{prevents any interference from happening}. The probability of detecting the photon in detector $B$ is now given by:
\begin{equation}\label{prob1}
P(A)=P(B)=1/2.
\end{equation}
Therefore, half of the time we detect the photon in detector $A$. The interaction with the environment has prevented us from determining that a superposition state was present -- but note that the combined state of the system and environment is still in a superposition state, we just can't confirm this. To introduce terminology that will be used later, we can say that the state of the environment is ``inaccessible'' to us. In this context, what we mean by this is that we cannot control and manipulate the environment. If we could control and manipulate it, we could ``erase'' the state of the environment, and this would allow the interference to be restored, and the superposition to be confirmed again. This idea of erasure is fascinating and highlights many important aspects of quantum mechanics, but a discussion of this is beyond the scope of this article (e.g. see the ``delayed choice quantum eraser'').

The above example is at the heart of decoherence: an inaccessible environment interacts with the system of interest, and prevents interference from taking place. With no interference, we cannot determine that a superposition state is/was present. While decoherence has not actually destroyed the superposition itself, it completely destroys our ability to \emph{confirm} the superposition state. If we cannot ever experimentally determine it, does this mean the superposition state does not exist? I'll return to this question later.

In situations with a system interacting with an inaccessible environment, it is often more convenient to calculate the final probabilities (e.g. equation~(\ref{prob1})) using the partial trace. If we only have access to part of a multi-partite system, then we can perform a partial trace over the inaccessible subsystems -- we ``trace them out''. For example, we can trace out the state of the environment in equation~(\ref{afterD}); it can be shown that this leaves the photon before the second beam splitter in the density matrix
\begin{align}
{1 \over 2} ( &\left|0 1\right\rangle\left\langle 0 1 \right|+\left|1 0\right\rangle\left\langle 1 0 \right| + \left\langle a | b \right\rangle (\left| 1 0 \right\rangle\left\langle 0 1\right| +\left| 0 1 \right\rangle\left\langle 1 0\right|) )\notag\\
&= {1 \over 2} ( \left|0 1\right\rangle\left\langle 0 1 \right|+\left|1 0\right\rangle\left\langle 1 0 \right| ). \label{posttr}
\end{align}
We then calculate the probabilities of detection to find exactly the same result as equation~(\ref{prob1}). With a little thought it becomes clear that the partial trace is just an alternative but equivalent method of calculating the final probabilities whenever we only have access to part of a state (the reason for this stems from the fact that $\Tr(\rho M) = \Tr_E \Tr_S (\rho M)$).

Note that the state in equation~(\ref{posttr}) above is not a superposition state, it just represents a photon that is either in the state $\left|0 1\right\rangle$, or the state $\left|1 0\right\rangle$, with $50\% $ probability of being each. This is equivalent to tossing a coin and not knowing the answer: you know that the state is heads or tails, but you do not know which. In fact, if we have no access to the environment, then there is no way to distinguish the state in equation~(\ref{posttr}) from a completely classical state in which we know the photon is $\left|0 1\right\rangle$ or $\left|1 0\right\rangle$ but don't know which. In the language of decoherence, we see that the interaction with the inaccessible environment has reduced our superposition state to a state that is entirely equivalent to a classical probabilistic mixture.

While decoherence destroys superposition states, there exists special states, known as ``pointer states'', that survive the interaction with the environment (these states belong to the ``pointer basis'' that was introduced earlier). As an example, if the state $\left|0 1\right\rangle $ interacts with the ``environment'' in Figure~\ref{intf}, then after tracing out the environment we have
\begin{equation}
\left| 0 1 \right\rangle \left\langle 0 1\right|,
\end{equation}
which is just the same state written as a density matrix. Despite the fact that the photon has interacted with the environment, the state is completely unchanged -- this state is immune to decoherence. The same results will be found with the state $\left|1 0\right\rangle $, which is also a pointer state. This is a crucial part of decoherence: pointer states survive the interaction with the environment, but superpositions of pointer states are reduced to mixtures.

Decoherence is often used to explain why superposition states are never found in the macroscopic world. To illustrate this, I now give a highly simplified real-world example, which should nonetheless serve to demonstrate how decoherence works. Again consider Schr\"{o}dinger's cat, which can be in two states, $\left| dead \right\rangle$ or $\left| alive \right\rangle$. To simplify this explanation we can assume that if the cat is alive then it is sitting up, whereas if it is dead it is lying down. Imagine first that the cat is alive. In principle, we could send a single photon into the box, aimed at the cat's head. Because the cat is sitting up, the photon will bounce off its head. But if the cat was dead, and therefore lying down, the photon would pass straight through to the other side of the box. This photon therefore contains information about the state of the cat, and as with the interferometer example above we can formalise the interaction between the cat and the photon as follows:
\begin{align}
&\left| alive \right\rangle \left| r \right\rangle	\rightarrow	\left| alive \right\rangle \left|a\right\rangle \notag\\
&\left| dead \right\rangle \left| r \right\rangle	\rightarrow	\left| dead \right\rangle \left| d \right\rangle.
\end{align}
Here $\left|r\right\rangle $  is the state of the photon before it hits the cat (``ready''), $\left|a\right\rangle $ is the photon that has bounced off the cat's head, and $\left|d\right\rangle $  is the photon that has passed through to the other side of the box (these photons contain information about whether the cat is alive or dead, respectively, hence the labelling $a $ and $d $).
If the cat is initially in a superposition of dead and alive then the following interaction takes place:
\begin{equation}
{1 \over \sqrt{2}} (\left| dead \right\rangle+\left| alive \right\rangle)\left| r \right\rangle		\rightarrow	{1 \over \sqrt{2}} ( \left| dead \right\rangle\left| d \right\rangle+\left| alive \right\rangle\left|a \right\rangle ).
\end{equation}
We can then perform the same analysis as above and trace out the state of the photon, which gives the final state of the cat as
\begin{equation}\label{19}
{1 \over 2} ( \left| dead \right\rangle\left\langle dead\right| +\left| alive \right\rangle\left\langle alive\right| ).
\end{equation}
The cat is now in a mixture of being dead \emph{or} alive. Note that we have assumed that the two photon states -- bouncing or not bouncing off the cat's head -- are completely orthogonal to one another. Their wave functions will have zero overlap for all practical purposes (FAPP), and therefore this approximation is valid. Note that as with the above example, the superposition hasn't actually been destroyed, but it is impossible to determine the superposition, so \emph{FAPP} the superposition has been destroyed, and a classical state remains.

This argument can be extended to the more realistic case of having a huge number of photons -- and air particles -- in the box with the cat. After interacting with the cat, the environment (photons and air particles) will contain information about the state of the cat, and we name the final state of these environmental states $E_{dead} $ and $E_{alive} $. The same conclusions as above will hold as long as the overlap $\left\langle E_{dead} |E_{alive}\right\rangle $ is zero, but this overlap will be far less than the overlap of the single photon. Also note that in the above example with a single-photon-environment it is at least conceivable that we can control and manipulate the photon, and therefore erase its state, restoring the possibility of the cat interfering with itself. But for the realistic environment of an Avocado's number of particles and photons the environment is completely beyond our control, and can legitimately be traced out, thereby preventing the survival of the superposition state $\left| dead \right\rangle+\left| alive \right\rangle$.

If we repeat the same analysis with an alive cat, then a very different outcome happens. After interacting with the photon, we trace out the state of the photon to give:
\begin{equation}
\left| alive \right\rangle \left\langle alive\right|.
\end{equation}
But this is just a pure state of an alive cat. This is because the states $\left| dead \right\rangle$ and $\left| alive \right\rangle$ are pointer states, and as introduced above these states are not ``destroyed'' by decoherence. Presumably, though it has not yet been proved, all of the objects around you are in pointer states. For example, the chair you sit on is in a pointer state: it continuously interacts with the environment -- largely photons -- but it remains in the same state. However, if you were to prepare the chair in a superposition of two different locations, you would soon find that the superposition reduces to a classical mixture.

Decoherence has been highly successful in modelling a large number of ``realistic'' scenarios. For example, if a dust particle (the system) is prepared in a superposition of different locations, and then allowed to interact with numerous air molecules (the environment), it has been showed that the superposition reduces to a mixture. Similarly, a neuron cannot survive for long in a superposition of firing and not firing: the interaction with surrounding molecules reduces the superposition to a mixture in an extremely short timescale \cite{tegmark2000importance}.
%[Include the following:?	larger systems will decohere faster. This is simply because a large system will generally interact more with its environment – more photons hit a human per second than a cat – and as we have seen even a single photon is enough to decohere a system. This is the core reason why we do not observe macroscopic objects in a superposition. For atom – scale systems, talented experimentalists have manage to reduce the interaction with the environment enough to keep the system in a superposition state for a time in the order of seconds. ]

We can now return to a key question addressed in this article: Why can we only measure everyday objects in a certain basis? Decoherence has not answered this question, but it does tell us why everyday objects are only stable when they are prepared in states in a certain basis -- the pointer basis -- and superpositions of these states are destroyed. In particular, decoherence does not allow us to conclude why our everyday world is objective. Only certain states might survive the interaction with the environment, but nothing we have seen so far prevents us from measuring these objects in a different basis. I could measure my cat in the basis $\{\ket{+}, \ket{-} \}$, and if I did this then I would obtain a completely different set of measurement results to someone else measuring in the basis $\{\ket{dead}, \ket{alive}\} $. Something more is needed to explain objectivity, and for this we turn to quantum Darwinism.

\subsection{Introduction to quantum Darwinism}
\begin{figure}[t!]
\centering
\includegraphics[trim=4cm 3cm 15cm 2cm, clip=true, width=7.5cm]{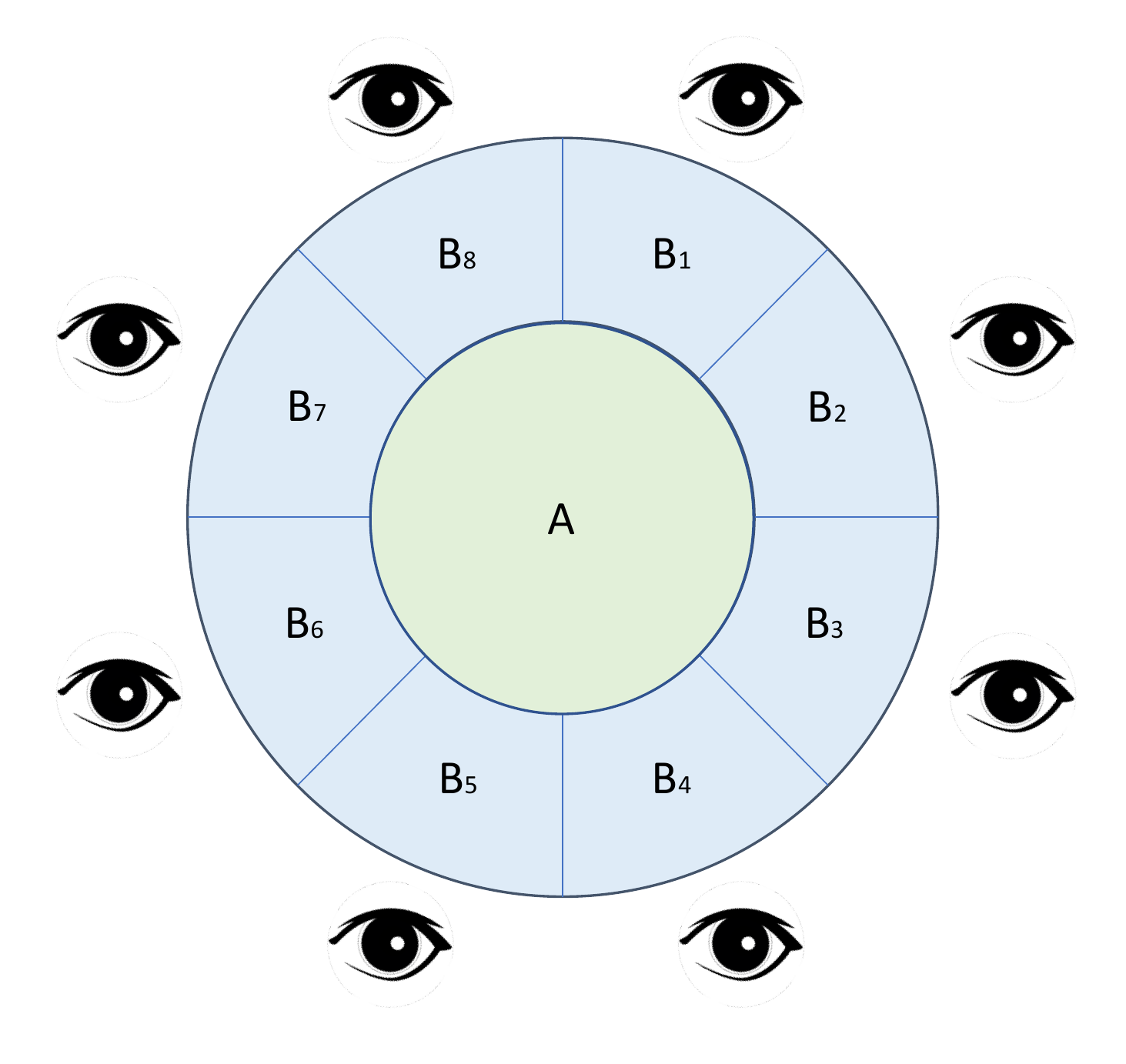}
\caption{In quantum Darwinism, the system $A$ interacts with an environment, but importantly the environment is now split into $N$ fragments. In the figure we have $8$ environments, labelled $B_1$ to $B_8$. Each observer, represented as an eye, has access to only one of the environment fragments.
\label{QD_quick}}
\end{figure}

In quantum Darwinism we still have a system and an environment, but now we say that the environment is divided into different fragments, as shown in Figure~\ref{QD_quick}. In this model, an observer wishing to gain information about the system does not measure the system directly, but rather probes one of the fragments of the environment. Despite perhaps seeming artificial at first, this model is closely linked to how we observe objects in our everyday world. Pick an object within your line of sight, and think about how you are gaining information about this object. When I observe my mug of tea, I do this by measuring the photons entering my eye, which had previously interacted with the tea. But the photons I measure are only a tiny fraction of the full photon environment, and the vast majority of the environment is completely inaccessible to me. Even with state-of-the-art experimental equipment it would be impossible to measure all of the photons that interact with the mug. Again we have an inaccessible environment, although unlike in decoherence now there is a small fraction that \emph{is} accessible to me. %In the example in Figure~\ref{artwork}, multiple artists are observing, and then painting, a cat. Again we can imagine splitting the photon environment into different fragments, and we can say that each artist only has access to one of the sections of the environment -- namely, those photons that enter their eyes. Here the relevant details about the cat –- its position, colour, shape, etc -- are objective, and each artist would paint the cat the same (that is, if we ignore artistic licence and perspective!).

%===========Fig:================
%====================================

As with decoherence, I will introduce quantum Darwinism using an oversimplified model, but this should give the flavour of how quantum Darwinism works. Take a system that can have one of two states, $\ket{\uparrow}$ or $\ket{\downarrow}$ (a qubit). Now imagine an environment containing $ N $ fragments, where each fragment can just have two states, $\left| 0 \right\rangle $ or $\left| 1 \right\rangle $. Clearly this environment is highly unrealistic, but it should help to give an intuitive introduction to quantum Darwinism. The environment is initially in the state: $\left| 000... 0 \right\rangle $, where $000... 0 $ represents $N $ zeros. In this example, the system-environment interaction is as follows:
\begin{align}\label{21}
&\ket {\uparrow, 000... 0 } \rightarrow \ket {\uparrow, 000... 0 } \notag\\
&\ket{\downarrow, 000... 0 } \rightarrow \ket{\downarrow, 111... 1 }
\end{align} 
We see that if the system is in the down state, then each environment fragment records this by changing to state $\left| 1 \right\rangle$, and similarly the up state is recorded by each environment remaining in state $\left| 0 \right\rangle $.
We first study the case when the system starts in a superposition state of up and down. The interaction with the environment will therefore be as follows:
\begin{align}
{1 \over \sqrt{2}} ( \ket {\uparrow, 000... 0 } &+ \ket{\downarrow, 000... 0 } ) \\ &\rightarrow {1 \over \sqrt{2}} ( \ket {\uparrow, 000... 0 } + \ket{\downarrow, 111... 1 } ).
\end{align} 
Now imagine you have access to one, and only one, of the environment fragments. As previously, we formulate this by tracing over all of the inaccessible environments (and the system, which is also inaccessible to us). Whichever environment you have access to, the final state of this environment after tracing out the others is given by:
\begin{equation}
{1 \over 2} \left( \left| 0 \right\rangle\left\langle 0 \right| +\left| 1\right\rangle\left\langle 1 \right| \right).
\end{equation}
Our environment is now in a mixture of zero and one; by measuring our environment, we can learn nothing about the fact that the system was initially in a superposition state -- all we can learn is that the system was in a state with equal weighting between up and down. But the initial state need not have been a superposition. It is straightforward to show that if the system started in the state $\propto \left| \uparrow \right\rangle\left\langle \uparrow\right| +\left| \downarrow \right\rangle\left\langle \downarrow \right| $, then after interacting with the environment, and tracing out all the inaccessible environments, our fragment would still be in the state $\propto \left| 0 \right\rangle\left\langle 0 \right| +\left| 1\right\rangle\left\langle 1 \right| $.

An important observation here is that every fragment of the environment \emph{contains the same information}. Therefore, the information about the system is objective: different observers with access to different parts of the environment will agree on their observations.

As with decoherence, it is instructive to also look at the case where the system starts in a state that is not a superposition, such as $\left| \uparrow \right\rangle$. Again considering the interaction in equation (\ref{21}) and focusing on one environment fragment (by tracing out the others), then regardless of which environment fragment you choose it will be in the state $\left| 0\right\rangle$. In this case, each environment fragment perfectly records the state of the system. The information is objective because each environment fragment contains the same information. Reconciling this with decoherence, we see that the states $\ket{\uparrow}$ and $\ket{\downarrow} $ are pointer states because they ``survive'' the interaction with the environment. This reveals why this model is termed quantum Darwinism.\\

\noindent 1) Certain states -- the pointer states -- survive the interaction with the environment. \\

\noindent 2) Information about these special states is proliferating into the environment. The states that survive \emph{copy the information} about themselves multiple times.\\

The connection with survival and reproduction in Darwinian natural selection is evident. But here there is only one ``generation'': it is not clear that the states of the environment themselves undergo a Darwinian process. For this reason, Charlie Bennett has instead termed this model quantum spam: certain states do survive the interaction, but it is a closer analogy to say that they \emph{spam their information} into multiple copies.

One way to think about this model is that the environment is performing a measurement on the system, and storing the results of that measurement. We saw that when the system was in the up state, each environment state stored this information (by remaining in the state $\left| 0\right\rangle$). If any observer then measures their environment fragment they can then extract this information. Now if the system is in a superposition of up and down, then if we perform a measurement on the state in the basis $\{\ket{\uparrow}, \ket{\downarrow}\} $, then we would obtain the result up with $50\% $ probability, and likewise for down. This is precisely the information contained in the environment after the interaction: each environment fragment is in the state $\propto \left| 0 \right\rangle\left\langle 0 \right| +\left| 1\right\rangle\left\langle 1 \right| $. \emph{The only information available} by measuring this environment fragment is information about a measurement in the $\{\ket{\uparrow}, \ket{\downarrow}\} $ basis. 

This point becomes clearer when we imagine measuring in a different basis, such as the $\{\ket{+},\ket{-}\} $ basis (defined as $\ket{\pm} \propto \ket{0} \pm \ket{1}$). If we had \emph{direct access to the system}, then a measurement in this basis would give the answer plus with $100\% $ probability. But if we only have access to an environment in a superposition of $0$ and $1$, then measuring this state in the $\{\ket{+},\ket{-}\} $ basis (where $\ket{+}$ and $\ket{-}$ are here defined in terms of $\ket{0}$ and $\ket{1}$) would just give plus with $50\%$ probability, and minus with $50\%$  probability. Information about the $\{\ket{+},\ket{-}\} $ basis has not survived the system-environment interaction. This gives us a crucial result of quantum Darwinism: \\

\hangindent=0.5cm \emph{Objectivity of observables: by probing a fragment of the environment, only information about a preferred basis is available. If we wish to gain information about a different basis, this is simply not possible.} \\

At least in the context of this very simple example, we have answered the question at the heart of this article: why is it that we seem to be forced to measure in a particular basis? By thinking about how we actually observe the world around us -- by probing a small fragment of the environment -- we have seen that \emph{only information about a particular basis is available}. We are not actually forced to measure in this basis, but if we chose to measure in a different basis then we would gain nonsensical information. Presumably, we have evolved to measure in the preferred basis. A hunter gatherer being pursued by a lion would not survive long if they tried to measure the lion in the superposition basis $\{here \pm there \} $! Unhelpful information would be obtained, and they would surely perish.

But of course it would be unjustifiable to take this simple model and extrapolate it into facts about objectivity in our everyday world. Quantum Darwinism has been studied in numerous models \cite{Blume2008,Zwolak2009,Riedel2012,Galve2015,Balaneskovic2015,Tuziemski2015,Tuziemski2015b,Tuziemski2016,Balaneskovic2016,Lampo2017,Pleasance2017} -- most of which are more realistic than my example above -- such as a photon environment interacting with a sphere \cite{Riedel2010}, or quantum Brownian motion \cite{Blume2008}. But so far these models are limited in scope, and due to the exponential growth of the Hilbert space required to model larger and larger systems, for the foreseeable future it will be impossible to simulate a macroscopic situation. Given this, how could we ever conclude that \emph{macroscopic} objectivity emerges from within quantum mechanics?

\subsection{The generic emergence of objectivity}
Constructing and studying larger and larger models \emph{is not the only way} to obtain results about macroscopic objects. Instead, if we can prove results about the mathematical structure of quantum mechanics itself; results of this nature would then apply to systems of all sizes. This is precisely what Brand{\~a}o, Piani, and Horodecki did in their influential paper, ``Generic emergence of classical features in quantum Darwinism'' \cite{brandao}. They showed that the \emph{objectivity of observables} is a generic feature of the mathematical structure of quantum mechanics.

To understand the results of \cite{brandao} (and a recent result by ourselves \cite{knott2018generic}), I must first give some background: I will introduce quantum channels (a.k.a quantum operation or cptp maps), and then the ``measure and prepare channel''. Quantum channels provide the most general way to describe how a quantum state $\rho_0$ evolves into a new quantum state $\rho_1$. The quantum channel $\Lambda $ acts as follows
\begin{equation}
\rho_0 \rightarrow \Lambda ( \rho_0 ) = \rho_1.
\end{equation}
Often in quantum mechanics we consider unitary evolution, which is given by
\begin{equation}
\rho_0 \rightarrow \Lambda ( \rho_0 ) = U \rho_0 U^{\dagger},
\end{equation}
for some unitary operator $U$. But quantum channels are more general than this, and cover situations in which the evolution is not unitary, for example in open quantum systems or when measurements are performed. As long as we begin in a quantum state, and end in a quantum state, then a quantum channel can be used to represent this map.

We can use a quantum channel $\Lambda$ to describe the evolution that takes place in the system-environment interaction shown in Figure~\ref{QD_quick}. The model we now consider is quite general. The system $A$ can be any state in any Hilbert space and similarly the environment $B$ can contain any environment state in any Hilbert space. For example, system $A$ could be the qubit described above with basis states $\{\left|\uparrow \right\rangle, \left|\downarrow \right\rangle\} $ and the environment can be qubits in $\{\left|0 \right\rangle, \left| 1 \right\rangle\} $; or the system $A$ could be a cup of tea, and the environment made of photons; or any other system-environment interaction you wish to study. We can divide the environment $B$ into as many different fractions as we require. In general, we have $n$ environment fragments, where each fragment is labelled $B_i$, for $i=1,..,n$, as shown in Figure~\ref{QD_quick}.

Again we assume that observers can only measure one fragment of the environment. To formalise this, the partial trace can also be given as a channel. In particular, the channel that traces out all environments except for environment $B_j$ is given by $\Tr_{\backslash B_j}$. We then give the full channel from the system $A$ to the environment $B_j$, after tracing out all other environments, as
\begin{equation}
\Lambda_j := \Tr_{\backslash B_j} \circ \Lambda
\end{equation}
where the symbol $\circ$ formalises how we combine channels (the channel on the right hand side always acts first). For example, $\Lambda_j$ could represent the quantum channel for how you observe a mug of tea: the photons interact with the tea, but the only fragment of the environment you have access to is $B_j$, and this whole interaction is formalised by the channel $\Lambda_j$. While it is near-impossible to write down exactly what this channel is, it is at least in principle possible, so it is legitimate to state that this channel is represented by $\Lambda_j$.

Next, I need to introduce the \emph{measure and prepare channel}, which is labelled $\cE_j$. Imagine the following task: you are given an ensemble of states
\begin{equation}
\left|\psi \right\rangle = \alpha \left|\uparrow \right\rangle + \beta \left|\downarrow \right\rangle
\end{equation}
where $\alpha$ and $\beta$ are (normalised) constants. You can only measure in the basis $\{\left|\uparrow \right\rangle, \left|\downarrow \right\rangle\} $. Your task is to tell your friend as much as you can about the ensemble of states. One possible strategy to do this is as follows: \\

\noindent 1) measure the ensemble of states in the basis $\{\left|\uparrow \right\rangle, \left|\downarrow \right\rangle\} $ \\

\noindent 2) your measurement outcomes allow you to determine the probability of up, $P(\uparrow)=|\alpha|^2$, and the probability of obtaining down $P(\downarrow)=|\beta|^2$ \\

\noindent 3) now you have gathered this information, you can encode it in a new state to send to your friend. For example, you could prepare the state
\begin{equation}
\rho = P(\uparrow) \sigma_1 + P(\downarrow) \sigma_2.
\end{equation}
Here $\sigma_1$ and $\sigma_2$ are arbitrary states of your choosing. For example, $\sigma_1$ could be a single photon state $\ket{1}$, and $\sigma_1$ could be a two photon state, $\ket{2}$. The important thing to note is that $\rho$ contains everything you could possibly know about the original state $\left|\psi \right\rangle$. This is because you can only perform measurements in the $\{\left|\uparrow \right\rangle, \left|\downarrow \right\rangle\} $ basis, and therefore the only information you can extract from the state is $P(\uparrow)$ and $P(\downarrow)$. The state $\rho$ contains this information, and therefore contains everything you could know about $\left|\psi \right\rangle$. \\

\noindent 4) Now send the state $\rho$ to your friend. Note that the state $\rho$ is just a classical mixture, and therefore you could send this information classically. For example, you could encode this information in a biased coin, which has probability of landing heads $P(\uparrow)$. You have now succeeded in your task: you have told your friend everything you know about $\left|\psi \right\rangle$. \\

We now wish to generalise this. We can replace the measurement in the basis $\{\left|\uparrow \right\rangle, \left|\downarrow \right\rangle\} $ with a generalised set of $m$ measurements $\{M_k\}$, where $k = 1... m$. Instead of the pure state $\left|\psi \right\rangle$, we can consider an arbitrary density matrix $\rho$. Finally, because we have $m$ measurements, we also need $m$ states $\sigma_{j,k}$ (the relevance of the subscript $j$ will become evident soon). Given this, we define the measure and prepare channel $\cE_j$ as
\begin{equation}\label{MP_channel}
\cE_j(\rho) := \sum_k \Tr (M_k X ) \sigma_{j.k}
\end{equation}
where $\rho$ is an arbitrary density matrix.

We are now ready to see a simplified version of the main result in \cite{brandao} -- readers interested in the full mathematical result are directed to \cite{brandao} (and also \cite{knott2018generic}). Take any quantum channel $\Lambda$ from a finite dimensional system $A$ to a system $B$ (which can have infinite dimensions). Then consider the channel $\Lambda_j$, introduced above, which maps to just one of the subsystems $B_j$. (A simplification of) the main result in \cite{brandao} is to show that there exists a set of measurements $\{M_k\}$ such that
\begin{equation}\label{generic}
\Lambda_j \rightarrow \cE_j	    \hspace{0.5cm}{as}\hspace{0.5cm}	n \rightarrow \infty
\end{equation}
where $\cE_j$ is given above. (In fact this only applies to most choices of $j$, but I ignore this detail in the explanation below.)
In words, this says that any quantum channel $\Lambda_j$ becomes indistinguishable from a measure and prepare channel $\cE_j$ in the limit of a large number of fractions of the environment, $n$. However, despite the way this is presented, we do not need $n$ to go to infinity, we justly need $n$ to be large \cite{brandao}.

So what exactly does this mean? The first important point is that this applies to \emph{any quantum channel}: regardless of the details of the interaction between the system and environment, and regardless of what constitutes your system and environment (assuming the system $A$ has finite dimensions), any quantum channel will become indistinguishable from $\cE_j$ in the limit. So what exactly is $\cE_j$? We can understand $\cE_j$ using the same four steps given above. Specifically in this case, a measurement is performed on the system with a specific set of measurements $\{M_k\}$. Then each environment fragment is prepared in a state that \emph{only contains information} about the measurements $\{M_k\}$. Therefore, if an observer wishes to learn about the system $A$ by probing the environment $B_j$, the only information available in environment $B_j$ is information about this particular set of measurements. The observer might want to learn about a completely different set of measurements, but this is simply not possible if they can only access $B_j$.

Now comes the most important observation: the set of measurements $\{M_k\}$ is independent of $j$. This means that, regardless of which environment you have access to, you can only learn about one-and-the-same set of measurement results. In other words, the observables that we can learn about are objective: \emph{objectivity of observables emerges generically from the basic mathematical structure of quantum mechanics.}

It is instructive to compare this result with our quantum Darwinism example introduced above, but with a simple extension. Imagine that the system is prepared in the state
\begin{equation} 
\left|\psi \right\rangle =\alpha \left|\uparrow\right\rangle +\beta \left|\downarrow\right\rangle.
\end{equation}
It can be shown that, if the system and environment interact as shown in equation (\ref{21}), then the final state of each fragment of the environment will be
\begin{equation}
|\alpha|^2 \left| 0 \right\rangle\left\langle 0 \right| + |\beta|^2 \left| 1\right\rangle\left\langle 1 \right|.
\end{equation}
This is a measure and prepare channel that has measured the system in the $\{\ket{\uparrow},\ket{\downarrow}\}$ basis, and then prepared a mixed state in the $\{\ket{0},\ket{1}\}$ basis. This example illustrates that in many cases the result in equation (\ref{generic}) is very pessimistic: in this example the quantum channel from the system to the environments becomes a measure and prepare channel \emph{regardless of the size of $n$}. A more accurate description of the main result in \cite{brandao} is that any channel becomes objective for large enough $n$, but many channels will be objective for much smaller values of $n$.

The result in \cite{brandao} helps explain why, in the everyday microscopic world, we are only able to perform measurements in certain bases. This is a consequence of the manner in which we measure our surroundings, namely that we measure a small fragment of a vast environment. As we have seen, each fragment of the environment only contains a limited and restricted amount of information. But we should not get too carried away here. There will be many examples in which the number of environment fragments $n$ is not large enough to conclude that $\Lambda_j$ is indistinguishable from $\cE_j$. Furthermore, as briefly mentioned above not all of the environment fragments $B_j$ becomes objective; and I should also mention that the $\sigma$'s in the measure and prepare channel may not be mutually orthogonal, so the information about the measurements $\{M_k\}$ might not even be extractable. Despite these caveats, the fact that objectivity of observables is built into the mathematical structure of quantum mechanics is a remarkable and in some ways refreshing finding.

One important restriction of the results in \cite{brandao} is that they only apply when the system $A$ is finite dimensional. But this severely limits us from concluding that objectivity of observables emerges in many realistic situation. In particular, the position basis, which is clearly a crucial part of how we measure our surroundings, is infinite dimensional. Furthermore, continuous variable systems, such as electromagnetic waves, live in infinite dimensions. A recent result by ourselves has overcome this finite dimensional restriction to prove that objectivity of observables still emerges from infinite dimensional systems \cite{knott2018generic}. But this cannot be proved in a completely unrestricted system, so we considered two physically motivated restrictions: firstly, finite energy systems -- this arguably includes all realistic systems of interest -- and then systems with an exponential energy cut-off (see \cite{knott2018generic} for the definition of this), which include Gaussian systems. In both cases we proved bounds to show that as the number of environment fragments grows large, objectivity of observables does indeed emerge. In \cite{brandao}, the bound on objectivity depended only on the system dimensions and the number of environments; in contrast, our bounds show an explicit dependence on the system's energy in the first case, and the strength of the exponential cut-off in the second.

%\textbf{[Replace ``measurement'' with ``observable'' throughout this section... Preferred basis? Preferred observables? Make it clear in what's written above. ???]}

\subsection{What is the role of the observer?}
In quantum mechanics, the observer is sometimes given an essential role in the theory, even at the fundamental level. But in decoherence and quantum Darwinism this is not the case. This is seen most easily in the context of the ``generic emergence'' results of the previous section, where it was shown that any quantum channel becomes indistinguishable from a measure and prepare channel. But in the measure and prepare channel it is the \emph{environment} that performs the measurement. Then the results of this measurement are stored in the state of environment. We can then ask the question: \textbf{if} an observer wishes to learn about the system by measuring the environment, what information can they gain? But the results hold regardless of whether we ask this question or not: only certain information is available in the environment, regardless of whether an observer is present to obtain this information. A world in which no conscious observers are present would still be an objective world!

%\textbf{[Give a very brief explanation that decoherence gives no special role for the observer? And the observer certainly does not need to be conscious]??}

\subsection{The measurement problem}
While it may not be immediately obvious, decoherence and quantum Darwinism can shed significant light on \emph{the measurement problem} in quantum mechanics. Numerous suggestions have been put forward to resolve the measurement problem, but despite the fantastically successful predictive power of quantum mechanics, no universally-accepted interpretation yet exists.

I will introduce the measurement problem by returning to Schr\"{o}dinger's cat. As described above, Schr\"{o}dinger's cat is prepared in a superposition of being dead and alive:
\begin{equation}
{1 \over \sqrt{2}} ( \left| dead \right\rangle +\left| alive \right\rangle ).
\end{equation}
The cat is in a sealed box, completely isolated from the rest of the world. What would happen if you are then told to open the box and measure the cat? Before you open the box, we will say that you are in a ``neutral'' state, which we denote $\left|neutral \right\rangle$. If you open the box and see that the cat is still alive, we say that you will be happy (assuming you like cats), and therefore the following transformation will take place
\begin{equation}\label{dead_alive}
\left| alive \right\rangle\left| neutral \right\rangle	\rightarrow	\left| alive \right\rangle\left| happy \right\rangle
\end{equation}
But if the cat is dead, you will be sad:
\begin{equation}
\left| dead \right\rangle\left| neutral \right\rangle	\rightarrow	\left| dead \right\rangle\left| sad \right\rangle.
\end{equation}
The linearity of quantum mechanics, and in particular the Schr\"{o}dinger equation, implies that the superposition state of the cat in equation (\ref{dead_alive}) will evolve as follows:
\begin{align}
{1 \over \sqrt{2}}  (\left| dead \right\rangle &+\left| alive \right\rangle) \left| neutral \right\rangle	\notag\\ &\rightarrow	{1 \over \sqrt{2}} ( \left| dead \right\rangle\left| sad \right\rangle + \left| alive \right\rangle\left| happy \right\rangle ).
\end{align}
Note that this is not controversial in itself: it is just a straightforward application of the Schr\"{o}dinger equation. But this suggests that you are now in a superposition of happy and sad! On the face of it this conclusion seems to be absurd: our intuition very clearly says that we cannot be in a superposition. How can we overcome this apparent disagreement between what we perceive, and what is predicted by applying the Schr\"{o}dinger equation? One common resolution is to introduce a collapse postulate into quantum mechanics. This postulate says that the state collapses into one of the two possibilities:
\begin{equation}
\left| dead \right\rangle\left| sad \right\rangle		\hspace{0.5cm}{or}\hspace{0.5cm}	   \left| alive \right\rangle\left| happy \right\rangle.
\end{equation}
In the early days of quantum mechanics, collapse was just a postulate, and no explanation was given of how collapse takes place, or what causes it. But this introduces many difficult questions at the heart of the measurement problem: What causes the collapse? It is usually assumed that a measurement causes collapse: but what is a measurement? Often it is said that a ``measuring device'', or even a conscious observer, is what causes the collapse. But if macroscopic objects are made of quantum particles, \emph{what is so special} about a measuring device or a conscious human observer to cause collapse?

Over the years various theories have been introduced to explain collapse with the hope of answering the above questions. Various mechanisms have been proposed: complexity causes collapse -- the more complex a system, the more likely it is to collapse \citep{GRW}; or consciousness itself causes collapse \cite{Wigner,London,Schro,Miranker,Bierman,Chalmers,KOBI,Squires,Stapp1,Stapp2,Germine,Gao,Altaisky, Goswami,Thaheld,Rosenblum}; or gravity causes collapse -- the larger the mass, the more likely collapse will occur \citep{Penrose}. These models can therefore explain why Schr\"{o}dinger's cat is never seen, or measured, as being in a superposition state. But despite the popularity of these theories, they are far from complete. They have never been experimentally confirmed -- while the \emph{consequences} of collapse are often evident, \emph{the collapse itself} has never been observed -- and furthermore collapse theories have not yet been extended to relativistic quantum mechanics, which would be essential for a complete theory. In addition, as I describe in \cite{me_CCC}, theories in which consciousness causes collapse can have some absurd consequences.

A completely different resolution to the measurement problem exists, which does not involve adding an extra collapse postulate or mechanism to quantum mechanics. Here the Schr\"{o}dinger equation itself, with nothing added or modified, is used to explain the appearance of collapse. Given the great success of the Schr\"{o}dinger equation, which itself is responsible for quantum mechanics often being referred to as ``our most successful theory ever'', this would be a desirable result. 

As with the rest of this article, the important point is that systems in the real world are never truly isolated, and always interact with an inaccessible environment. Assuming there are photons in the box with the cat, then as I explain below equation (\ref{19}), before opening the box the cat and the photon environment are actually in the state
\begin{equation}
{1 \over \sqrt{2}} (\left| dead \right\rangle\left| E_{dead}\right\rangle + \left| alive \right\rangle\left| E_{alive}\right\rangle )
\end{equation}
where $E_{dead}$ and $E_{alive}$ represent the state of the environment after it has interacted with a dead cat, or an alive cat, respectively. We then open the box, giving the state
\begin{equation}
{1 \over \sqrt{2}} ( \left| dead \right\rangle\left| sad \right\rangle\left| E_{dead} \right\rangle + \left| alive \right\rangle\left| happy \right\rangle\left| E_{alive} \right\rangle	)		%(X2)	
\end{equation}
The final state is now an entangled state, and we are now in a superposition of being happy (having observed the alive cat), and sad (having observed the dead cat). But how can we \emph{confirm} this superposition? Interference is the only way to confirm this superposition, but as I showed below equation (\ref{19}) the inaccessible environment prevents us from ever confirming this superposition. If we could fully control all of the photons in the environment, then in principle it would be possible to ``erase'' the information they contain, thereby allowing interference to take place. But clearly we cannot do this: the photons remain inaccessible, and we remain unable to confirm the superposition. This is formalised by tracing out the environment, which gives
\begin{align}\label{X1}
{1 \over 2} ( \left| dead \right \rangle\left| sad \right\rangle &\left\langle dead \right| \left\langle sad \right|  \notag\\
&+ \left| alive \right\rangle\left| happy \right\rangle \left\langle alive \right| \left\langle happy \right| )			%(X1)
\end{align}
Thus, for all intents and purposes, the final state is a mixture: as far as we can tell we are either happy (having observed the alive cat), or sad (having observed the dead cat). No measurement we could ever realistically perform could tell us otherwise. This explains the \emph{appearance} of collapse, and indeed the final state is the same as if we had invoked a collapse postulate. But here we have used the Schr\"{o}dinger equation alone to demonstrate why and how ``collapse'' happens.

Note that the arguments used up to this point should still not be seen as controversial: it would be hard to argue that there are no photons in the box, or that the photons do not interact with the cat; and if the photons interact with the cat it is clear that their trajectories will contain information about the state of the cat; and the different environmental states of the photons (corresponding to different macroscopic states of the cat) will clearly be orthogonal; and finally the state of the environment will be inaccessible to us.

But the state in equation (\ref{X1}) is not the full picture -- this is just the perceived state, given that we can't access the environment. The real state is
\begin{equation}\label{X2}
{1 \over \sqrt{2}} (\left| dead \right\rangle\left| sad \right\rangle\left| E_{dead} \right\rangle + \left| alive \right\rangle\left| happy \right\rangle\left| E_{alive} \right\rangle		)	%(X2)	
\end{equation}
How should we interpret this? Most scientists, through most of history, implicitly assume some kind of realism when interpreting science: the world around us really does exist, independent of our own existence, and we can perform experiments in order to learn more about this world. If we follow this reasoning and interpret the state in equation (\ref{X2}) as a real physical system, then despite the fact that there is no (practical) measurement we could perform to confirm it, \emph{we are actually in a superposition state!} This is, in essence, the Everett interpretation of quantum mechanics: We take the Schr\"{o}dinger equation, and the Schr\"{o}dinger equation alone, and use this to explain the appearance of collapse. Collapse is no longer a mysterious and controversial postulate, it is just a \emph{prediction} of the Schr\"{o}dinger equation whenever an inaccessible environment is involved. A realist interpretation of the final state then says that a \emph{macroscopic system has entered into a superposition state}. This is the idea of the ``many worlds'' that emerge in the Everett interpretation: the different parts of the superposition can never interfere or interact with one another, so for all intents and purposes they can be considered as separate ``worlds'', although really they are just different parts of the same superposition state.

The ``branching'' or ``splitting'' of the wave function into superpositions of distinct macroscopic states happens whenever a measurement is performed on a superposition state. But what is meant by ``measurement'' here? All of the things we normally consider to be a measurement -- such as measuring the spin of a particle, or measuring the state of Schr\"{o}dinger's cat by peering into the box -- have some important features in common. These measurements accurately record the state of the system (e.g. ``the spin is up'' or ``the cat is alive''), and furthermore they record this information in macroscopically distinct states (e.g. a computer screen displaying ``up'', or the state of the person's brain who has measured the cat). These macroscopic states will always interact with an inaccessible and orthogonal environment, and therefore we always need to trace out the environment, resulting in a mixed state of the combined system-measuring apparatus. As soon as the system and measuring apparatus interact a macroscopic superposition is created; and we can say that the branching happens at the point when the environment contains enough information so that the different parts of the macroscopic superposition can never realistically interfere.

This brief discussion of the Everett interpretation certainly leaves many questions unanswered, but this is not the right place to give a complete and thorough description of the theory. The reader is referred to Tegmark for a non-technical introduction \cite{Tegmark}, and Wallace for a thorough description \cite{Wallace}. Instead, our motivation here is to argue that the Everett interpretation should not be seen as a radical interpretation of quantum mechanics. It just results from a straightforward application of the Schr\"{o}dinger equation, followed by realist interpretation of the final state. Of course the conclusion -- that there are many ``copies'' of you simultaneously going about your day in different ``branches'' of a larger superposition state -- is it quite hard to swallow. But it is important to realise that this is \emph{not a postulate} of quantum mechanics, but rather a \emph{prediction}. In particular, the Everett interpretation certainly does not postulate the creation of whole new universes every time a measurement is performed.

Where does this leave collapse theories? It is hard to believe that any of the steps leading up to equation (\ref{X1}) can be denied, given that they just rely on a straightforward application of quantum mechanics. So, at least in this example, \emph{the appearance of collapse} occurs without needing to add additional dynamics to the theory, as is done in collapse theories. In this sense, collapse theories are certainly not \emph{necessary} in order to explain collapse: given that it is for all practical purposes impossible to confirm that you are in a superposition such as that in equation (\ref{X2}), collapse theories have no added explanatory value. But they come at the cost of having to modify the Schr\"{o}dinger equation -- arguably the most successful equation in physics. 

It should be noted that the example above is just one specific example, and decoherence has certainly not yet \emph{completely} explained why we observe the macroscopic world the way we do. To fully model the world would require an impossibly large quantum computer. However, results such as \cite{brandao} and \cite{knott2018generic} show that it is possible to explain certain important aspects of our reality -- namely the objectivity of observables -- in a model independent way. The Holy Grail of this endeavour would be to explain all aspects of our classical reality using only the Schr\"{o}dinger equation, supplemented by a few indisputable assumptions.

I should mention that the realist interpretation discussed above is not the only way to interpret equation (\ref{X2}). For example, one could argue that the wavefunction in quantum mechanics represents our knowledge of the system, rather a real physical system itself (e.g see QBism \cite{Fuchs,Mermin}). Many more interpretations exist, but this is not the place to give a full introduction to all the competing interpretations.

%\textbf{CAPTION:? We finish this section by introducing figure X [Joe's illustration]. In section X we saw that any quantum channel $\Lambda$ becomes objective (in the sense of objectivity of observables) when the number of environment fragments is large enough -- this is depicted by all of the painters painting the cat identically. But an objective, classical cat is not the full state of the system: the real state is $\Lambda (\rho)$, which in general will be a highly entangled and unimaginably complicated state containing within it numerous macroscopically distinct ``branches''!}

\subsection{Quantum Darwinism and the Everett interpretation}

Above we saw that decoherence uses unaltered quantum mechanics (a.k.a. unitary quantum mechanics) to explain why we never see macroscopic objects in a superposition; and that if we are to give a realist interpretation to quantum mechanics, then decoherence suggests that the wavefunction forms branches that contain different macroscopic states. In the Everett interpretation these branches have some distinct features: The branches must be orthogonal to one another, as otherwise there would be some interference between the different branches -- i.e. the dead cat would interfere with the alive cat. And if there are multiple observers on a branch, then all the observers should agree with one another, as otherwise reality would not be objective, which would be in direct conflict with our everyday experiences. Furthermore, macroscopic objects in our surroundings must exist in accordance with our observations of them. For example, your chair might be in different locations on different branches, but when you sit on your chair you are sure that your observations of it -- by measuring photons in the environment -- are in agreement with the actual location of the chair in your branch.

There is still much work to be done on proving (or indeed disproving) that unaltered quantum mechanics does indeed lead to this description of a branching reality. But already quantum Darwinism has come a long way in this regard, for example \cite{riedel2017classical} (although it is interesting to note that papers on quantum Darwinism are rarely explicit about any connection with the Everett interpretation, perhaps because this is off-putting to many readers). In this article I have explained how a certain important aspect of classicality, objectivity of observables, emerges generically, in any system-environment interaction, so long as there are enough environment fragments. A natural question then arises: is it possible to prove that Everett-style branches emerge generically in quantum theory?

One reason why the results in \cite{brandao} and \cite{knott2018generic} fall short of answering this question is that, while the final state of each environment becomes indistinguishable from the mixture $\sum_k \Tr (M_k \rho ) \sigma_{j.k}$ (see equation (\ref{MP_channel})), the states $\sigma_{j.k}$ are not orthogonal to one another, as would be required for this to represent a branched state. An interesting extension of \cite{brandao,knott2018generic} would therefore be to investigate under what conditions the states $\sigma_{j.k}$ do become mutually orthogonal. This might not be possible with a completely arbitrary quantum channel, but perhaps it can be done by placing some reasonable restrictions on the channel. With this in mind we may ask: what general properties are shared by all every-day channels? It would be absurd to consider whether an object in a pitch-black room is objective; all relevant channels therefore must involve some minimum amount of information transfer from the system to all relevant environment fragments. An alternative restriction could be to coarse-grain the measurements the observers are allowed to perform (or e.g. follow the strategy in \cite{ mironowicz2017monitoring}).

\subsection{Conclusion}

If we apply the quantum formalism to isolated systems, then we can find any number of seemingly absurd predictions, most notably that macroscopic objects such as tables, chairs, cats, and even humans can enter into a superposition state. This clearly violates our picture of reality, and at first sight it seems impossible that many aspects of our classical reality, such as objectivity, can be explained with unaltered quantum mechanics alone. However, in hindsight this line of reasoning has a serious flaw: systems in the real world are \emph{never} isolated from their environment -- macroscopic objects are continuously interacting with photons, air particles, and all the other objects they come into either direct or indirect contact with. This was the great insight of decoherence, and by now it is well known that a system interacting with its environment cannot remain in a superposition state for long. This reasoning alone cannot explain objectivity, but by elevating the environment to a channel of communication between a system and multiple observers, we have seen that objectivity can also emerge.

Despite these developments, the majority of results in quantum Darwinism were obtained using specific models. But it is challenging enough to construct a realistic model of a molecule, never mind a macroscopic object such as a cat. For this reason, the method of studying specific models might never allow us to draw complete conclusions about the macroscopic world, and a number of important questions regarding the quantum-to-classical transition will remain unanswered. To overcome this, we saw here that Brand{\~a}o et al. \cite{brandao} and ourselves \cite{knott2018generic} have instead looked at the mathematical structure of quantum mechanics, and this method allows us to draw completely general conclusions, and so infer things about the macroscopic world itself. We saw that objectivity is built into the mathematical structure of quantum theory, and so it is completely consistent that macroscopic objects have objective properties, despite the fact that they are made of quantum mechanical particles. The question of how we can reconcile quantum physics with an understanding of our everyday world has challenged scientists for a century, but this ongoing work in decoherence and quantum Darwinsim is now providing a solution. \\

\begin{acknowledgments}
P.K. acknowledges discussions with Gerardo Adesso, Marco Piani, and Tommaso Tufarelli. This work was supported by the Foundational Questions Institute (fqxi.org) under the Physics of the Observer Programme (Grant No.~FQXi-RFP-1601).
\end{acknowledgments}

%=============Bibliography============
\bibliographystyle{apsrevfixedwithtitles}
\bibliography{QD_inf_plus_essay}
\end{document}